\documentclass[12pt]{article}
 \usepackage{amsmath,amsfonts,latexsym,amssymb,amsthm}
 \usepackage{times}

   \newcommand{\field}[1]{\mathbb{#1}}
   \newcommand{\rz}{\field{R}}
   \newcommand{\cz}{\field{C}}
   \newcommand{\zz}{\field{Z}}
   
   \newcommand{\dd}[1]{\frac{\partial}{\partial #1}}
   \newcommand{\dzd}[1]{\frac{\partial^2}{\partial #1^2}}
   \newcommand{\dzdz}[2]{\frac{\partial^2}{\partial #1 \partial #2}}

   \newcommand{\dirac}{/\!\!\!\nabla}
   
   \newcommand{\daggern}{/\!\!\!n}

   \newcommand{\openset}{\mathcal{O}}
   \newcommand{\Hilbert}{\mathcal{H}}
   \newcommand{\feld}{\mathcal{F}}

 \newtheorem{theorem}{Theorem}[section]
 \newtheorem{definition}[theorem]{Definition}
 
 \newtheorem{lem}[theorem]{Lemma}
 \newtheorem{cor}[theorem]{Corollary}
 \newtheorem{pro}[theorem]{Proposition}
 
 \newtheorem{rem}[theorem]{Remark}

\begin{document}

\title{The Reeh-Schlieder Property for Quantum Fields on Stationary Spacetimes}

\author{Alexander Strohmaier}

\date{\small Universit\"at Leipzig,
Institut f\"ur theoretische Physik,
Augustusplatz 10/11, D-04109 Leipzig, Germany\\
E-mail: alexander.strohmaier@itp.uni-leipzig.de}
\maketitle
\noindent
\begin{abstract}
  \noindent
  We show that as soon as a linear quantum field on a stationary spacetime
  satisfies a certain type of hyperbolic equation, the (quasifree) ground- and KMS-states
  with respect to the canonical time flow have the Reeh-Schlieder property.
  We also obtain an analog of Borchers' timelike tube theorem.
  The class of fields we consider contains the Dirac field, the Klein-Gordon
  field and the Proca field.
\end{abstract}

{\small \bf Mathematics Subject Classification (2000):} 81T05, 81T20, 35J15, 35L10 \\

\maketitle

\section{Introduction}

For the analysis of quantum field theory in curved spacetime it has
turned out that the framework of algebraic quantum field theory
(see \cite{Haag:1992hx}) is most suitable for analyzing
the problems connected with the non-uniqueness of the quantization of linear
fields (see e.g. \cite{Wald} and the references therein).
The problem reduces to finding appropriate representations of the field
algebra which can straightforwardly be constructed on manifolds (see \cite{Dimock:1980hf,Dimock:1982hf}).
This is the same as specifying vacuum-like states over the algebra.
On stationary spacetimes it is possible to distinguish states
as ground- or KMS-states with respect to the canonical time translations.
For free fields satisfying certain wave-equations it was recently shown in \cite{Verch:vs00}
that such passive quasifree states satisfy the microlocal spectrum condition (see
\cite{Brunetti:1996rf,Radzikowski:1992eg}) which is believed to be a
substitute for the usual spectrum condition in Minkowski spacetime.
For the case of the scalar field on a \mbox{4-dimensional} globally hyperbolic
spacetime it was shown in \cite{Radzikowski:1992eg} that
the microlocal spectrum condition is equivalent to the requirement that the 2-point
function is of Hadamard form (see e.g. \cite{Fulling:1989nb, Wald}), which
allows for a renormalization of the stress energy tensor
(\cite{Wald:1977up}).
It therefore seems reasonable to consider ground- or KMS-states for free
quantum fields on stationary spacetimes as good substitutes
for the vacuum in flat spacetime.\\
In the Minkowski space theory the vacuum vector turns out to be cyclic for field
algebras associated to nonvoid open regions (\cite{Reeh:1961re}).
This Reeh-Schlieder property of the vacuum vector holds also 
for thermal quantum field theories (\cite{Jakel:1999ji}).
By the lack of symmetry in general
spacetimes it is not clear whether physically reasonable vacuum states have
this property as well. It was shown in \cite{Verch:1993pn} that the
quasifree ground-state of the massive scalar field on an ultrastatic spacetime is of this
kind.
The proof uses an anti-locality property of the square root
of the Laplace operator.
Using similar methods it was possible to obtain a Reeh-Schlieder-type property for solutions to the
Dirac equation on an ultrastatic spacetime with compact Cauchy surface
in \cite{Baer:99}, and the Reeh-Schlieder property for ground- and KMS-states of
the free Dirac field on a static globally hyperbolic 4-dimensional
spacetime in \cite{stroh:99}.\\
We will show in this paper that the Reeh-Schlieder property of ground- and KMS-
states holds for a large class of free fields on stationary spacetimes.
We introduce the notions of (classical) linear fermionic and bosonic field
theories on a spacetime and show that these notions
lead via canonical quantization to quantum field theories
on this spacetime. We note that our approach does not use an initial data formulation.
In the last section we show how the most common free fields
fit into this framework.
Our main theorem states that as soon as the classical
field fulfills a certain hyperbolic partial differential equation,
a state over the field algebra of the quantized theory, which
is a (quasifree in the bosonic case) ground- or KMS-state with respect to the group of time
translations, has the Reeh-Schlieder property. In order to show this we
combine the Gelfand-Maurin theorem on generalized eigenvectors with classical results
on the strong unique continuation property of solutions of certain second order elliptic
differential equations. As a result we obtain a curved spacetime analog to the timelike tube
theorem of Borchers (\cite{Borchersr}). Using standard arguments this yields the
Reeh-Schlieder property.\\
The class of fields which fulfill our assumptions contains the Dirac field, the
Proca field and the scalar field on arbitrary connected
4-dimensional globally hyperbolic stationary Lorentzian manifolds.
We note that as a consequence the Hartle-Hawking state for the Klein-Gordon
field on the external Schwarzschild spacetime
has the Reeh-Schlieder property (see \cite{Kay:1985zs}).
This is a new result which so far has not been obtained by the methods
previously employed.\\
The Reeh-Schlieder property serves as the starting point for the use of the
Tomita-Takesaki-theory within quantum field theory. The application of this theory
to the analysis of quantum field theory in Minkowski spacetime was very
fruitful (see e.g. \cite{Borch:99})
and there might be an impact as well on curved space quantum
physics (see \cite{Buchholz:1998pv},\cite{Guido:1999xu}).

\section{Classical linear field theories}

In the following and throughout the text a smooth manifold
will always be Hausdorff and separable as a topological space.
If we are given a smooth vector bundle $E$ over a smooth manifold $M$ one can endow
the space of smooth sections $\Gamma(E)$ and the space of compactly
supported smooth sections $\Gamma_0(E)$ with locally convex
topologies (see \cite{Dieud3,Dieud7}) in a similar way as for
$C^\infty(\rz^n)$ and $C^\infty_0(\rz^n)$.
These locally convex vector spaces turn out to be nuclear
(see \cite{Maurin:1968}, or \cite{Pietsch:1965,Jarchow} for properties of nuclear
spaces).
We denote the topological dual of the space $\Gamma_0(E)$ by
$\mathcal{D}'(M,E^*)$, calling it the space of distributions with values in the dual
bundle $E^*$.
For some open $\openset \subset M$ the subspaces $\Gamma_0(E,\openset)$
and $\Gamma(E,\openset)$ of sections supported in $\openset$ are closed.
We call a smooth vector bundle $E$ a $G$-vector bundle for some Lie-group $G$ if
there is a smooth action of $G$ on the base space $M$ and on $E$,
such that the bundle projection $E \to M$ is equivariant and the left action
$q: E_x \to E_{qx}$ is linear for all $q \in G$ and $x \in M$.
In this case we have a canonical action of $G$ on $\Gamma_0(E)$ and
$\Gamma(E)$, for which we use the notation
$G \times \Gamma(E) \to \Gamma(E),\; (q,f) \to qf$.
These representations of $G$ are continuous in the corresponding
locally convex topologies.
A smooth n-dimensional Lorentzian manifold $(M,g)$ is a smooth n-dimensional
manifold with smooth metric $g$ of constant signature $(1,-1,\ldots,-1)$ and
$n \geq 2$ (see e.g. \cite{Neill:1983}).
For any open set $\openset$ we denote
the open causal complement of $\openset$, i.e. the interior of the set of points
in $M$ which cannot be joined to a point in $\openset$ by
a causal curve, by $\openset^\perp$.
If $E$ is a smooth vector bundle over a Lorentzian manifold $(M,g)$
we say a second order differential operator $P: \Gamma(E) \to \Gamma(E)$
has metric principal part, if in local coordinates
the principal part is of the form $\mathbf{1} \cdot g^{ik} \partial_i\partial_k$.

\subsection{The fermionic case}

Let $(M,g)$ be a smooth n-dimensional Lorentzian manifold. We denote the identity component
of the group of isometries of $M$ by $G$ and its universal covering group by
$\tilde G$.
Note that $\tilde G$ is a Lie-group which acts smoothly on $M$.

\begin{definition} \label{Fermi}
  A linear fermionic field theory on $M$ is a 5-tuple
  $(\Hilbert,\mathcal{K},E,\eta, \rho)$, where
  $\Hilbert$ is a complex Hilbert space with conjugation $\mathcal{K}$, $E$ a smooth complex
  $\tilde G$-vector bundle, $\eta$ a linear map
  $\eta:\Gamma_0(E) \to\Hilbert$ with dense range, and $\rho$ a
  unitary representation of $\tilde G$ on $\Hilbert$ commuting with $\mathcal{K}$,
  such that the following conditions are satisfied:
  \begin{enumerate}
    \item Covariance:
      $\eta(q f) = \rho(q)(\eta(f)), \quad \forall f \in \Gamma_0(E), q \in \tilde G$.
    \item Causality: $\openset_1 \subset \openset_2^\perp$ implies 
      $\Hilbert(\openset_1) \perp \Hilbert(\openset_2)$, where
      $\Hilbert(\openset)$ denotes the closure of the image of $\Gamma_0(E,\openset)$ under
      $\eta$.
    \item Continuity: $\eta$ is weakly continuous, i.e.
      $\langle v, \eta(\cdot) \rangle$ is in
      $\mathcal{D}'(M,E^*)$ for all $v \in \Hilbert$.
    \item $\mathcal{K}$ is local, i.e. $\mathcal{K}(\Hilbert(\openset))=\Hilbert(\openset)$.
  \end{enumerate}
\end{definition}

\begin{pro} \label{Fermicont}
 For any linear fermionic field theory $(\Hilbert,\mathcal{K},E,\eta, \rho)$
 on $M$ the map $\eta$ is norm continuous. Moreover, the
 representation $\rho$ is strongly continuous.
\end{pro}
\begin{proof}
  The sesquilinear form $B(f,g):=\langle \eta(f),\eta(g) \rangle$ on
  $\Gamma_0(E) \times \Gamma_0(E)$ is separately continuous, and as a consequence of the
  nuclearity of $\Gamma_0(E)$ it is jointly continuous. This gives
  the norm continuity of $\eta$.
  Since the action of $\tilde G$ on $E$ is smooth, the action
  of $\tilde G$ on $\Gamma_0(E)$ is continuous.
  As a consequence we obtain the strong continuity of $\rho$.
\end{proof}

\subsection{The bosonic case}

Let $(M,g)$ be again a smooth n-dimensional Lorentzian manifold and
$G$ be the identity component of the group of isometries
of $M$.

\begin{definition} \label{Bose}
  A linear bosonic field theory on M is a 5-tuple
  $(\mathcal{W},\sigma,E,\eta, \rho)$, where
  $\mathcal{W}$ is a symplectic vector space with symplectic form
  $\sigma$, $E$ a smooth real $G$-vector bundle, $\eta$ a surjective linear map
  $\eta:\Gamma_0(E) \to\mathcal{W}$, and $\rho$ a
  representation of $G$ on $\mathcal{W}$ by symplectomorphisms,
  such that the following conditions are satisfied:
  \begin{enumerate}
    \item Covariance:
      $\eta(q f) = \rho(q)(\eta(f)), \quad \forall f \in \Gamma_0(E), q \in G$.
    \item Causality: $\openset_1 \subset \openset_2^\perp$ implies 
      $\mathcal{W}(\openset_1) \perp \mathcal{W}(\openset_2)$, where
      $\mathcal{W}(\openset)$ denotes the image of $\Gamma_0(E,\openset)$ under $\eta$.
    \item Continuity: For all $v \in \mathcal{W}$ the map
      $\Gamma_0(E) \to \rz, \; f \to \sigma(v, \eta(f) )$ is continuous, i.e. 
      defines a distribution in $\mathcal{D}'(M,E^*)$.
  \end{enumerate}
\end{definition}

\section{Canonical Quantization}

In this section $(M,g)$ will be a smooth n-dimensional Lorentzian manifold.
We will work in the framework of algebraic quantum field theory
(\cite{Haag:1992hx}). A quantum field theory on $M$ will be defined by a net of
local field algebras, i.e. a map $\openset \to \feld(\openset)$
from the relatively compact open subsets of $M$ to the set of closed $*$-subalgebras
of a $C^*$- or $W^*$-algebra $\feld$ which is isotone, i.e.
$\feld(\openset_1) \subset \feld(\openset_2)$ whenever $\openset_1 \subset \openset_2$.

\subsection{Quantization of a linear fermionic field theory}

We show how a linear fermionic field theory gives rise to a quantum
field theory on $M$ given by a net of local field algebras.\\
Given a linear fermionic field theory $(\Hilbert,\mathcal{K},E,\eta, \rho)$,
the field algebra $\mathcal{F}$ is the (self-dual) CAR-algebra
$\textrm{CAR}(\mathcal{H}, \mathcal{K})$ (see \cite{Arak:1971ar}). 
This is the $C^*$-algebra with unit generated by symbols $B(v)$ with
$v \in\mathcal{H}$ and the relations
\begin{gather} \label{carrel}
v \to B(v) \quad \textrm{is complex linear,}\\
B(v)^* = B(\mathcal{K} v),\\
\lbrace B(v_1) , B(v_2) \rbrace = B(v_1)B(v_2)+B(v_2)B(v_1)=\langle \mathcal{K} v_1,v_2 \rangle\;.
\end{gather}
$\feld$ has a natural $\zz_2$-grading. The even/odd parts are spanned by those
products $B(v_1) \ldots B(v_k)$  with an even/odd number of generators. 
For each relatively compact subset $\mathcal{O} \subset M$ we define the
local algebra $\mathcal{F}(\mathcal{O}) \subset \mathcal{F}$ to be the
closed unital $*$-subalgebra generated by the symbols
$B(\eta(f))$ with $f \in \Gamma_0(E,\openset)$.
The representation $\rho$ of the group $\tilde G$ on $\Hilbert$ gives rise to
a representation $\tau$ of $\tilde G$ by strongly continuous Bogoliubov
automorphisms of the algebra (see \cite{Arak:1971ar}). It is not difficult to
check the following properties of the net
$\openset \to\mathcal{F}(\openset)$:
\begin{enumerate}
  \item Isotony: $\openset_1 \subset \openset_2$ implies
    $\mathcal{F}(\openset_1) \subset \mathcal{F}(\openset_2)$.
  \item Causality: if $\openset_1 \subset \openset_2^\perp$, then
        $\lbrace \feld(\openset_1),\feld(\openset_2) \rbrace=\lbrace 0 \rbrace$,\\
        where $\lbrace \cdot,\cdot \rbrace$ denotes the graded commutator. 
  \item Covariance:
    $\tau(q) \mathcal{F}(\openset)=\mathcal{F}(q \openset) \quad \forall q \in \tilde G$.
\end{enumerate}
Moreover, $\feld$ is the quasilocal algebra of the net
$\openset \to\mathcal{F}(\openset)$ (see \cite{Bratteli2}, proposition 5.2.6).
Hence, this defines a reasonable quantum field theory on the manifold $M$.
Note that $\feld$ is not the algebra of observables.
The algebra of observables $\mathcal{A}$ should be a *-subalgebra
of $\feld_{\textrm{even}}$ consisting of elements $a$ for which
$\tau_{g_1} a=\tau_{g_2} a$ whenever $p(g_1)=p(g_2)$, where
$p: \tilde G \to G$ is the covering map. Usually $\mathcal{A}$
is the $*$-subalgebra of elements which are
invariant under the action of a gauge group.

\subsection{Quantization of a linear bosonic field theory}

Each linear bosonic field theory $(\mathcal{W},\sigma,E,\eta, \rho)$
gives rise to a quantum field theory
on $M$.
The field algebra $\feld$ is defined to be the CCR-algebra
$\textrm{CCR}(\mathcal{W},\sigma)$ (see \cite{Kay:1993gr,Kay:1991mu,Bratteli2}).
This is the $C^*$-algebra generated by symbols $W(v)$ with
$v \in\mathcal{W}$ and the relations
\begin{gather} \label{ccrrel}
  W(-v)=W(v)^*,\\
  W(v_1)W(v_2) = e^{-i \sigma(v_1,v_2)/2}W(v_1+v_2).
\end{gather}
We define for each relatively compact open subset $\openset \subset M$
the local field algebra $\feld(\openset) \subset \feld$ to be the closed $*$-subalgebra
generated by the symbols $W(v)$ with $v \in \mathcal{W}(\openset)$.\\
The representation $\rho$ of $G$ gives rise to a representation $\tau$
of $G$ by Bogoliubov automorphisms of the algebra $\feld$ (see e.g.\cite{Bratteli2}), and the net
$\openset \to \feld(\openset)$ has the following properties:
\begin{enumerate}
  \item Isotony: $\openset_1 \subset \openset_2$ implies
    $\mathcal{F}(\openset_1) \subset \mathcal{F}(\openset_2)$.
  \item Causality: if $\openset_1 \subset \openset_2^\perp$, then
        $[\feld(\openset_1),\feld(\openset_2)]=\lbrace 0 \rbrace$. 
  \item Covariance: $\tau(q) \mathcal{F}(\openset)=\mathcal{F}(q \openset) \quad \forall q \in G$. 
\end{enumerate}
Moreover, $\feld$ is the quasilocal algebra of the net
$\openset \to\mathcal{F}(\openset)$ (see \cite{Bratteli2}, proposition 5.2.10).
Hence, this defines a reasonable quantum field theory on the manifold $M$. 
Unlike the fermionic case the representation $\tau$ fails to be strongly
continuous whenever it is nontrivial.
We therefore need to pass to certain representations of the field algebra
to obtain a net of von Neumann algebras on which $\tau$ extends to a
$\sigma$-weakly-continuous representation.
In order to avoid complications we specialize to the so-called quasifree
states.
Assume that we are given a scalar product $\mu$ on $\mathcal{W}$
which dominates $\sigma$, i.e. satisfies the estimate
\begin{gather} \label{dom}
 \vert \sigma(v_1,v_2) \vert^2 \leq 4 \mu(v_1,v_1) \mu(v_2,v_2) \quad
 v_1,v_2 \in \mathcal{W}.
\end{gather}
In this case the linear functional $\omega_\mu : \feld \to \cz$, defined by
\begin{gather}
  \omega_\mu(W(v)):=e^{-\mu(v,v)/2} \quad v \in \mathcal{W},
\end{gather}
is a state (see \cite{Kay:1993gr,Kay:1991mu,Bratteli2}). The states over $\feld$ which can be realized in this
way are called quasifree states.
A quasifree state $\omega_\mu$ gives rise to a one particle structure (Proposition 3.1 in \cite{Kay:1991mu}), that
is a map $K_\mu: \mathcal{W} \to H_\mu$ to some complex Hilbert space $H_\mu$, such
that
\begin{enumerate}
  \item the complexified range of $K_\mu$, (i.e. $K_\mu \mathcal{W} + i K_\mu \mathcal{W}$), is
        dense in $H_\mu$,

  \item $\langle K_\mu v_1,K_\mu v_2 \rangle=\mu(v_1,v_2)+\frac{i}{2}\sigma(v_1,v_2)$.
\end{enumerate}
This structure is unique up to equivalence.
A one particle structure $(K_\mu,H_\mu)$ for a quasifree state allows one to
construct the GNS-triple
$(\pi_{\omega_\mu},\Hilbert_{\omega_\mu},\Omega_{\omega_\mu})$ explicitly
(see \cite{Kay:1991mu,Kay:1993gr,Bratteli2}).
Namely, one takes $\Hilbert_{\omega_\mu}$ to be the bosonic Fock
space over $H_\mu$ with Fock vacuum $\Omega_{\omega_\mu}$, and defines 
$\pi_{\omega_\mu}(W(v))=\textrm{exp}(-(\overline{\hat a^*(Kv)-\hat a(K v)}))$,
where $\hat a^*(\cdot)$ and $\hat a(\cdot)$ are the usual creation and
annihilation operators.
One clearly has the following
\begin{pro} \label{Bosedens}
 Let $\omega_\mu$ be a quasifree state over the CCR-algebra
 $\feld=\textrm{CCR}(\mathcal{W},\sigma)$ and let
 $(\pi_{\omega_\mu},\Hilbert_{\omega_\mu},\Omega_{\omega_\mu})$ be its
 GNS-triple.
 If $V \subset \mathcal{W}$ is a subspace which is dense in $\mathcal{W}$
 in the topology defined by $\mu$, then the $*$-algebra generated by the set
 \begin{gather*}
   \lbrace \pi_{\omega_\mu}(W(v)), v \in V \rbrace \subset \pi_{\omega_\mu}(\feld)
 \end{gather*}
 is strongly dense in the von Neumann algebra $\pi_{\omega_\mu}(\feld)''$.
\end{pro}

\begin{definition}
 Let $\feld$ be a field algebra constructed from a
 linear bosonic field theory $(\mathcal{W},\sigma,E,\eta, \rho)$
 on $M$. We call a quasifree state $\omega_\mu$ over $\feld$ continuous
 if the map
 \begin{gather*}
  \Gamma_0(E) \times \Gamma_0(E) \to \rz, \quad (f_1,f_2) \to \mu(\eta(f_1), \eta(f_2))
 \end{gather*}
 is continuous and hence defines a distribution in
 $\mathcal{D}'(M \times M, E^* \boxtimes E^*)$, where
 $E^* \boxtimes E^*$ is the direct product of the bundle $E^*$ with itself
 over the base space $M \times M$.
\end{definition}
\noindent
This is clearly a necessary and sufficient condition for Wightman 2-point function
$w_2(\cdot,\cdot):=\langle K \eta(\cdot), K \eta(\cdot) \rangle$
to be a distribution.
For the class of continuous quasifree states we can circumvent the problems connected with
the non-continuity of the representation $\tau$ of $G$ on $\feld$.
Since the action of $G$ is continuous on $\Gamma_0(E)$ and $\rho$
leaves $\sigma$ and $\mu$ invariant, there exists a unique
strongly continuous representation $\tilde U$ of $G$ on the one particle
Hilbert space $H_\mu$, such that $K_\mu \circ \rho(q) = \tilde U(q) \circ K_\mu$ for all
$q \in G$. Second quantization gives a strongly continuous unitary
representation $U$ of $G$ on $\Hilbert_{\omega_\mu}$, such that
$\pi_{\omega_\mu}(\rho(q) a)=U(q) \pi_{\omega_\mu}(a) U^{-1}(q)$
for all $q \in G$ and $a \in \feld$.
Hence, one gets the following proposition:

\begin{pro} \label{Bosecont}
  Let $\omega_\mu$ be a continuous $G$-invariant quasifree state over the field algebra
  $\feld$ constructed from a
  linear bosonic field theory $(\mathcal{W},\sigma,E,\eta, \rho)$ on $M$.\\
  Let $(\pi_{\omega_\mu},\Hilbert_{\omega_\mu},\Omega_{\omega_\mu})$ be its
  GNS-triple and $U$ be the unitary representation of $G$ on
  $\Hilbert_{\omega_\mu}$ induced by $\tau$.
  Then $U$ is strongly continuous and hence $\tau$ can be continued
  to a $\sigma$-weakly-continuous representation $\hat\tau$ of
  $G$ by automorphisms of the von Neumann algebra
  $\hat\feld:=\pi_{\omega_\mu}(\feld)''$.
\end{pro}
\noindent
Therefore, given a continuous quasifree state $\omega_\mu$ we can construct the net
of von Neumann algebras
$\openset \to \hat\feld(\openset):=\pi_{\omega_\mu}(\feld(\openset))''$.
This assignment is isotone, causal and covariant, and the representation $\hat \tau$ of $G$
is $\sigma$-weakly-continuous. It gives rise to a quantum
field theory on $M$ with reasonable physical properties.

\section{The Reeh-Schlieder property for quantized linear fields}

\subsection{Stationary spacetimes}

Let $(M,g)$ be an n-dimensional time-oriented Lorentzian manifold which admits a one parameter group
$h_t$ of isometries, smooth in $t$, with timelike orbits giving rise to a timelike Killing
vector field $\xi$. Such a manifold is called stationary.
For later considerations we need a special class of charts.
\begin{lem}\label{pos}
 For each point $p \in M$ there exists an open neighbourhood $\openset$
 and a chart $\phi:  \openset \to \rz^n$ with coordinates
 $(x_0,\ldots,x_{n-1})$, such that in local coordinates
 \begin{enumerate}
   \item $\xi = \dd{x_0}$,
   \item the $(n-1) \times (n-1)$ matrix
         $-g^{\alpha \beta}(x),\; \alpha,\beta=1,\ldots,n-1$
         is positive for all $x \in \phi(\openset)$.
 \end{enumerate}
\end{lem}

\begin{proof}
 One can always choose a neighbourhood $\openset_1$ of $p$
 and a chart ${\phi:  \openset_1 \to \rz^n}$ with coordinates
 $(x_0,\ldots,x_{n-1})$, such that $\xi = \dd{x_0}$ and the dual metric tensor is diagonal
 in the point $p$, i.e.
 \begin{gather}
  -g^{ik}(\phi(p))=\textrm{diag}_n(g(\xi,\xi),-1,\ldots,-1).
 \end{gather}
 $-g^{\alpha \beta}(\phi(p)),\; \alpha,\beta=1,\ldots,n-1$ is then a positive
 matrix. Since the matrix-valued function $-g^{\alpha \beta}$ is continuous,
 there exists a neighbourhood $\phi(\openset)$ of $\phi(p)$, on which it is positive. 
\end{proof}

\subsection{Free quantum fields on stationary spacetimes}

Since $h_t$ is a group of isometries it defines a group homomorphism
$\rz \to G$ which lifts uniquely to a group homomorphism $\rz \to \tilde G$.
Hence, in case we have a net of field algebra $\openset \to \feld(\openset)$ constructed from a linear
fermionic or bosonic field theory we canonically get a one parameter group
of automorphisms $\tau_t$ which acts covariantly, i.e.
$\tau_t \feld(\openset)=\feld(h_t \openset)$.
We call this group the group of canonical time translations induced by $h_t$.
In the interesting cases one can realize $\feld$ on a Hilbert space, such that
$\tau_t$ is $\sigma$-weakly-continuous and hence extends to an automorphism
group $\hat\tau_t$ of the von Neumann algebra $\feld''$ (see e.g. proposition \ref{Bosecont}). 
It is then possible to distinguish
vacuum states over the field algebra $\feld$ as ground- or KMS-states with respect
to the group of time translations $\tau_t$.
\begin{definition}
Let $\mathcal{A}$ be a $W^*$-algebra and $\alpha_t$
be a $\sigma$-weakly-continuous one-parameter group of
$*$-automorphisms of $\mathcal{A}$.
An $\alpha_t$ invariant normal state $\omega$ is called ground-state
with respect to $\alpha_t$ if the generator of the corresponding
strongly continuous unitary group $U(t)$ on the GNS-Hilbert space is positive.
\end{definition}
\begin{definition}
Let $\mathcal{A}$ and $\alpha_t$ be as above.
A normal state $\omega$ is called KMS-state with inverse temperature $\beta>0$
with respect to $\alpha_t$ if for any pair $A,B \in \mathcal{A}$
there exists a complex function $F_{A,B}$ which is analytic in the strip
$$\mathcal{D}_\beta:=\lbrace z \in \cz; 0<Im(z)<\beta \rbrace$$ and bounded
and continuous on $\overline{\mathcal{D}_\beta}$, such that
\begin{gather*}
 F_{A,B}(t)=\omega(A \alpha_t(B)),\\
 F_{A,B}(t+i\beta)=\omega(\alpha_t(B)A)\;.
\end{gather*}
\end{definition}

\begin{definition}
 Let $(M,g,h_t)$ be a connected  stationary Lorentzian manifold.
 Let $\lbrace \mathcal{F}(\mathcal{O}) \rbrace$ be the net of local field
 algebras constructed from a linear fermionic field theory
 $(\Hilbert,\mathcal{K},E,\eta, \rho)$ or from a linear bosonic field theory
 $(\mathcal{W},\sigma,E,\eta, \rho)$.
 Denote the group of canonical time translations by $\tau_t$.
 Let $\omega$ be a $\tau_t$-invariant state
 over the quasilocal algebra $\feld$, which we assume to be quasifree
 and continuous in the bosonic case, and denote by
 $(\pi_\omega,\Hilbert_\omega,\Omega_\omega)$ the corresponding GNS-triple.
 We say that $\omega$ is a ground or KMS state, if the unique normal extension
 of $\omega$ over $\pi_\omega(\feld)''$ is a ground or KMS state with respect
 to the unique $\sigma$-weakly-continuous extension of
 the group of time translations $\tau_t$.
\end{definition}
\noindent
In the fermionic case the existence of ground states
is always guaranteed (\cite{Arak:1971ar}).
Moreover, there exists a unique quasifree KMS-state
with inverse temperature $\beta >0$ (see \cite{Bratteli2,Arak:1971ar}).
In the bosonic case the construction of continuous
quasifree ground- and KMS-states seems problematic
in the general case. See e.g. \cite{Kay:1978yp}
for conditions that allow the construction of a continuous quasifree
ground state for the Klein-Gordon quantum field on a 4-dimensional
stationary spacetime.

\subsection{A density theorem and a tube theorem}

\begin{theorem} \label{hyperth}
 Let $(M,g,h_t)$ be a connected stationary Lorentzian manifold and
 $E$ a smooth complex $h_t$-vector bundle.
 Let $H$ be a complex Hilbert space
 and $\rho_t$ a strongly continuous unitary one-parameter group on $H$. 
 Assume that we have a linear strongly continuous map
 $\hat\eta: \Gamma_0(E) \to H$ with dense range which is covariant, i.e.
 $\hat\eta(h_t f)=\rho_t \hat\eta(f)$ for all $t \in \rz, f \in \Gamma_0(E)$.
 Assume furthermore that  $\hat\eta \circ P =0$ for some
 second order differential operator $P$ with metric principal part.
 Then $\hat\eta( \Gamma_0(E,h_\rz \openset))$ is dense in $H$ for each nonvoid
 open set $\openset \subset M$.
\end{theorem}
\begin{proof}
We introduce the following notations:
\begin{gather}
  V:=\hat\eta( \Gamma_0(E,h_\rz \openset)^\perp,\\
  p_V \ldots \; \textrm{orthogonal projection onto} \; V,\\
 \Phi:= \textrm{Ran}(p_V \circ\hat\eta).
\end{gather}
$V$ is clearly a $\rho_t$-invariant subspace and $\Phi$ is dense
in $V$. Identifying $\Phi$ with
$\Gamma_0(E)/\textrm{ker}(p_V \circ\hat\eta)$
we can endow $\Phi$ with the locally convex quotient topology.
Since $\Gamma_0(E)$ is nuclear and $p_V \circ\hat\eta$
is continuous, $\Phi$ is a nuclear space (see \cite{Maurin:1968,Pietsch:1965,Jarchow}), and clearly the
inclusion map $\Phi \to V$ is continuous.
We denote the dual of $\Phi$  by $\Phi'$.
It follows that
\begin{gather*}
  \Phi \subset V \subset \Phi'
\end{gather*}
is a Gelfand triple.
We denote the selfadjoint generator of the group $\rho_t \vert_V$ by $A$.
Clearly, $\Phi \subset \mathcal{D}(A)$ and $A$ restricts to a continuous operator $\Phi \to \Phi$.
Moreover $\Phi$ is invariant under the action of $\rho_t \vert_V$. As a consequence $A$ is essentially
selfadjoint on $\Phi$.
Hence, there exists a complete set of generalized eigenvectors (see
\cite{Gelfand:1964,Maurin:1968}) for $A$, i.e. a family
$v_\lambda \in \Phi'$ indexed by a subset $I \subset \rz$,
such that
\begin{gather}
 v_\lambda( A f ) = \lambda v_\lambda(  f ),
 \quad \forall f \in \Phi,\\
 v_\lambda (f)=0\; \textrm{for all } \lambda \in I \Leftrightarrow  f=0. 
\end{gather}
By continuity
\begin{gather}
 \psi_\lambda(\cdot):=v_\lambda( p_V \circ \hat\eta(\cdot) )
\end{gather}
defines for each $\lambda \in I$ a distribution in $\mathcal{D}'(M,E^*)$, such that
\begin{gather}
 \psi_\lambda(\Gamma_0(E,h_\rz \openset))=\lbrace 0 \rbrace,\\\label{elie}
 \psi_\lambda(\mathcal{L}_\xi f) = i \lambda \psi_\lambda(f),\\
 \psi_\lambda(P \; \cdot) = 0,
\end{gather}
where $\xi$ is the timelike Killing vector field induced by $h_t$
and $\mathcal{L}_\xi$ the Lie derivative on $\Gamma_0(E)$ defined by
$\mathcal{L}_\xi f=\lim_{t \to 0} \frac{h_t f -f}{t}$.\\
For each point $p \in M$ there exists an open contractible neighbourhood $\mathcal{U}$
and a chart mapping $\mathcal{U}$ to $\rz^n$ which we can choose to be of the form constructed
in lemma \ref{pos}.
The restriction of $E$ to $\mathcal{U}$
is trivial and we can identify $\Gamma_0(E,\mathcal{U})$ with
$C^\infty(\mathcal{U}) \otimes \cz^N$ in such a way that
$\mathcal{L}_\xi f = \dd{x_0} f$ for both functions and sections.
We consider the distribution $\psi_\lambda$ in such a chart.\\
We have $P^* \psi_\lambda=0$, where $P^*$ is the adjoint operator.
Moreover, equation (\ref{elie}) reads ${\dd{x_0}\psi_\lambda=-i \lambda \psi_\lambda}$.
Note that the principal part of $P^*$ has the form
\begin{gather}
 g^{00}\dzd{x_0} + g^{0\alpha} \dzdz{x_0}{x_\alpha}+
 g^{\alpha \beta}\dzdz{x_\alpha}{x_\beta}\\
 \quad \alpha,\beta = 1,2,\ldots, n-1. \nonumber
\end{gather}
Replacing
the $x_0$-derivatives by $-i \lambda$ and adding the term
$-\dzd{x_0} - \lambda^2$ we obtain an elliptic second order
differential operator $P_e$ with $P_e \psi_\lambda=0$.
Hence, $\psi_\lambda$ is smooth (see
e.g.\cite{HormBook:1990,Dencker:1981de})
and since $P_e$ has scalar principal part the 
classical result of Arozajn \cite{Aroz:1957ar} (see especially \mbox{Remark 3})\footnote{
For a detailed treatment see section 17.1 of \cite{Hormbook:1985} and the
references therein} implies that $\psi_\lambda=0$ in each such chart in which
$\psi_\lambda$ vanishes in an open nonvoid set, in particular in each such chart
intersecting with $h_\rz \openset$. Since $M$ is connected this
implies $\psi_\lambda=0$ on $M$.
The set of generalized eigenvectors $v_\lambda$ was complete and therefore
$V$ equals $\lbrace 0 \rbrace$ and the theorem is proved.
\end{proof}
\noindent
As a consequence one gets a result similar to the timelike tube theorem
(\cite{Borchersr}) in Minkowski spacetime.
\begin{theorem}\label{Roehrensatz}
 Let $(M,g,h_t)$ be a connected  stationary Lorentzian manifold.
 Let $\lbrace \mathcal{F}(\mathcal{O}) \rbrace$ be a net of local field
 algebras constructed from a linear fermionic field theory
 $(\Hilbert,\mathcal{K},E,\eta, \rho)$ or from a linear bosonic field theory
 $(\mathcal{W},\sigma,E,\eta, \rho)$.
 Assume that $\eta \circ P=0$ for some second order differential operator $P$
 with metric principal part.
 One has
 \begin{enumerate}
  \item \bf The fermionic case. \it 
    The $*$-subalgebra of $\feld$ generated by the subset $\;$
    $\bigcup_{t \in \rz} \feld(h_t \openset)$ is norm dense in $\feld$
    for each nonvoid relatively compact open set $\openset \subset M$.
  \item \bf The bosonic case \it 
    Let $\omega_\mu$ be a quasifree and continuous state over the quasilocal
    algebra $\feld$ .
    Assume that $\omega_\mu$ is invariant under the automorphism group
    $\tau_t$ induced by the Killing flow $h_t$. Denote its GNS-triple by
    $(\pi_{\omega_\mu}, \mathcal{H}_{\omega_\mu},\Omega_{\omega_\mu})$.
    The $*$-subalgebra of $\pi_{\omega_\mu}(\feld)$ generated by the subset $\;$  
    $\bigcup_{t \in \rz} \pi_{\omega_\mu}(\feld(h_t \openset))$ is strongly dense in
    $\pi_{\omega_\mu}(\feld)$ for each nonvoid relatively compact open set $\openset \subset M$.
 \end{enumerate}
\end{theorem}

\begin{proof}
We start with the fermionic case.
 Let $\Hilbert_1$ be the subspace of $\Hilbert$ generated by the set
 \begin{gather*}
  \lbrace \eta(h_t f) \in \Hilbert; f \in \Gamma_0(E,\mathcal{O}), t \in \rz \rbrace,
 \end{gather*}
 and let $\feld_1$ be the unital $*$-subalgebra of $\feld$ generated by the set
 $\lbrace B(v), v \in \Hilbert_1 \rbrace$.
 Clearly, $\feld_1$ is equal to the $*$-subalgebra of $\feld$ generated
 by $\bigcup_{t \in \rz} \feld(h_t \openset)$.
 The group of time translations $\rho_t$ is a strongly continuous
 one parameter group on $\Hilbert$, and with $\hat\eta=\eta$
 we can apply theorem \ref{hyperth} above. It follows that $\Hilbert_1$ is dense in $\Hilbert$ and hence
 $\feld_1$ is norm dense in $\feld$.\\
In the bosonic case let $\mathcal{W}_\cz$ be the complexification
of $\mathcal{W}$ and take the complexification of $\mu$
as a scalar product on $\mathcal{W}_\cz$. We complete this space
and obtain a Hilbert space $H$. We complexify the real vector bundle $E$ and
obtain the complex vector bundle $E_\cz \cong E \oplus E$ with a canonical action of $h_t$.
We can extend the map $\eta$ to a map $\eta_\cz$ which maps from the
section of $E_\cz$ to $H$. By construction $\eta_\cz$
has dense range.
Since $\mu$ is invariant under the action of the time translations $\rho_t$
we get a strongly continuous unitary action of the group of time translations
on $H$ such that $\eta_\cz$ is equivariant.
 Let $H_1$ be the complex subspace of $H$ generated by the set
 \begin{gather*}
  \lbrace \eta_\cz(h_t f) \in H; f \in \Gamma_0(E_\cz,\mathcal{O}), t \in \rz \rbrace.
 \end{gather*}
 We see that all the assumptions for theorem \ref{hyperth} are fulfilled and hence
 $H_1$ is dense in $H$. Using proposition \ref{Bosedens}
 one therefore concludes that the $*$-subalgebra of $\pi_{\omega_\mu}(\feld)$
 generated by the set
 $\lbrace W(\eta(h_t f)),  f \in \Gamma_0(E,\mathcal{O}), t \in \rz \rbrace$
 is strongly dense in $\pi_{\omega_\mu}(\feld)$.
\end{proof}

\subsection{The Reeh-Schlieder property for ground- and KMS-states}

\begin{definition}
 Let $\lbrace \mathcal{F}(\mathcal{O}) \rbrace$ be a net of local
 field algebras indexed by the relatively compact open subsets of a manifold $M$.
 Let $\omega$ be a state over the quasilocal field algebra $\mathcal{F}$
 and $(\pi_\omega, \mathcal{H}_\omega,\Omega_\omega)$ its GNS-triple.
 We say that $\omega$ has the Reeh-Schlieder property if $\Omega_\omega$
 is cyclic for the von Neumann algebra $\pi_\omega(\mathcal{F}(\mathcal{O}))''$ for
 each nonvoid relatively compact open set $\mathcal{O} \subset M$.
\end{definition}
\noindent
Our main theorem is:

\begin{theorem}\label{mainth}
 Let $(M,g,h_t)$ be a connected  stationary Lorentzian manifold.
 Let $\lbrace \mathcal{F}(\mathcal{O}) \rbrace$ be the net of local field
 algebras constructed from a linear fermionic field theory
 $(\Hilbert,\mathcal{K},E,\eta, \rho)$ or from a linear bosonic field theory
 $(\mathcal{W},\sigma,E,\eta, \rho)$.
 Assume that $\eta \circ P=0$
 for some second order differential operator $P$ with metric principal part.
 Let $\omega$ be a state over the quasilocal algebra $\feld$ which we assume to be quasifree and
 continuous in the bosonic case.
 If $\omega$ is a ground- or KMS-state with respect to the automorphism group
 $\tau_t$ induced by the Killing flow $h_t$, then $\omega$ has the Reeh-Schlieder property.
\end{theorem}
\noindent
We postpone the proof for a moment.
Let $\mathcal{O} \subset M$ be a nonvoid relatively compact open set and
$\mathcal{B}(\mathcal{O}) \subset \mathcal{F}(\mathcal{O})$ the
$*$-subalgebra consisting of those elements $a \in \mathcal{F}(\mathcal{O})$
for which there exists a neighbourhood $I \subset \rz$ of $0$, such that
$\tau_I(a) \subset \mathcal{F}(\mathcal{O})$. Let $\mathcal{O}_1 \subset M$ be another nonvoid
open set such that $\overline{\mathcal{O}_1} \subset \mathcal{O}$. We clearly
have the inclusions
\begin{gather}\label{incl}
 \mathcal{F}(\mathcal{O}_1) \subset \mathcal{B}(\mathcal{O})
 \subset \mathcal{F}(\mathcal{O})\;.
\end{gather}
We denote the strongly continuous unitary
group, implementing $\tau_t$ on the GNS-Hilbert space, by $\tilde U(t)$.
One has
\begin{lem}\label{invG}
 Let
 $\mathcal{E}:=\overline{\pi_\omega(\mathcal{B}(\mathcal{O})) \Omega_\omega}$. 
 Then $\mathcal{E}$ is invariant under the action of $\tilde U(t)$, i.e.
 $\tilde U(\rz) \mathcal{E} \subset \mathcal{E}$.
\end{lem}  
\begin{proof}
Let $\psi \in \mathcal{E}^\perp$. For each
$a \in \pi_\omega(\mathcal{B}(\mathcal{O}))$ we then
have at least for some open neighbourhood $I \subset \rz$ of 0
$$
 f(t):=\langle \psi, \tilde U(t) a\Omega_\omega \rangle=0 \quad \forall t \in I\;.
$$
Since $\omega$ is a KMS-state ($\beta>0$) or a ground-state ($\beta=\infty$),
$f(t)$ is the boundary value of a function $F(z)$ which is analytic
on the strip $$\mathcal{D}_{\beta/2}=\lbrace z \in \cz; 0<Im(z)<\beta/2
\rbrace$$ and bounded and continuous on $\overline{\mathcal{D}_{\beta/2}}$.
By the Schwartz reflection principle $f(t)$ vanishes on the whole real axis.
Therefore $\langle \tilde U(t) \psi, a\Omega_\omega \rangle=0$ for all
$t \in \rz$.
Hence, $\mathcal{E}^\perp$ is invariant under the action of $\tilde U(t)$.
\end{proof}
We are now able to give the proof of the main theorem:\vspace{0.3cm}\\
\it Proof of theorem \ref{mainth}. \rm\\
 We denote by $\mathcal{R}$ the von Neumann algebra
 $\bigvee_{t \in \rz} \tilde U(t) \pi_\omega(\mathcal{F}(\mathcal{O}_1)) \tilde U(t)^*$.
 Lemma \ref{invG} implies that $\mathcal{E}$ is invariant under the action of
 $\mathcal{R}$, i.e. $\mathcal{R}\mathcal{E} \subset \mathcal{E}$.
 By theorem \ref{Roehrensatz} $\mathcal{R}=\pi_\omega(\feld)''$.
 Hence $\Omega_\omega$ is cyclic for $\mathcal{R}$ and
 therefore $\mathcal{E}=\Hilbert_\omega$.
 By the inclusions (\ref{incl}) we have
 $\mathcal{E} \subset \overline{\pi_\omega(\feld(\openset)) \Omega_\omega}$
 and hence $\Omega_\omega$  is cyclic for $\pi_\omega(\feld(\openset))$.
\endproof

\section{Examples of linear field theories}

In this section $(M,g)$ will be an oriented time-oriented 4-dimensional Lorentzian
manifold which is globally hyperbolic in the sense that it admits a smooth
Cauchy surface. For some subset $\openset \subset M$ the set of points, which can be reached
by future/past directed causal curves emanating from  $\mathcal{O}$, will
be denoted by $J^\pm(\mathcal{O})$.\vspace{0.3cm}\\
\bf The free Dirac field \rm (see \cite{Dimock:1982hf}). 
 It is known that $M$ possesses a trivial spin structure
 given by a $\textrm{Spin}^+(3,1)$-principal bundle $SM$ and a two-fold covering
 map $SM \to FM$ onto the bundle $FM$ of oriented time-oriented orthonormal
 frames. One can now construct the Dirac bundle $DM$ which is associated to
 $SM$ by the spinor representation and is a natural module for the Clifford
 algebra bundle Cliff($TM$) (see \cite{Baum} and \cite{Lawson}).
 Furthermore, the Levi-Civita connection on $TM$
 induces a connection on $DM$ with covariant derivative
 $$\nabla : \Gamma(SM) \to \Gamma(SM \otimes T^*M).$$
 Given a vector field $n$ we write as usual $\daggern$ for the section in the
 Clifford algebra bundle $\gamma(n)$, or in local coordinates $\gamma^i n_i$. 
 There exists an antilinear bijection
 $\Gamma(DM) \to \Gamma(DM^*): u \to u^+$, the Dirac conjugation, which
 in the standard representation in a local orthonormal spin frame has the form
 $u^+=\overline{u} \gamma^0$, where the bar denotes complex conjugation in
 the dual frame.
 We use the symbol ${}^+$ also for the inverse map.
 Canonically associated with the Dirac bundle there is the Dirac operator
 which in a frame takes the form
 $$ \dirac = \gamma^i \nabla_{e_i}\;.$$
 The Dirac equation for mass $m \geq 0$ is
 \begin{equation}
  (-i \dirac + m) u = 0, \quad u \in \Gamma(DM)\;.
 \end{equation}
 The Dirac equation has unique advanced and retarded fundamental solutions\\
 $S^\pm: \Gamma_0(DM) \to \Gamma(DM)$
 satisfying
 $$ (-i \dirac + m) S^\pm = S^\pm (-i \dirac + m) = \mathrm{id} \quad \textrm{on}
 \quad \Gamma_0(DM)\;,$$
 $$ \textrm{supp}(S^\pm f) \subset J^\pm(\textrm{supp}(f))\;.$$
 We can define the operator $S:=S^+-S^-$ and form
 the pre-Hilbert space
 \begin{gather}
  H:=\Gamma_0(DM)/\textrm{ker}(S)
 \end{gather}
 with inner product
 \begin{gather}\label{fip}
 \langle [u_1], [u_2] \rangle_H := -i \int_M (u_1^+, S u_2)(x) d\mu(x),
 \end{gather}
 where $(\cdot,\cdot)$ denotes the dual pairing between the fibres of $DM$
 and $DM^*$ and $\mu(x)$ the canonical (pseudo)-Riemannian measure on $M$.
 $[f]$ denotes the equivalence class containing $f$.
 Analogously we form the pre-Hilbert space
 \begin{gather}
   H^+ :=\Gamma_0(DM^*)/(\textrm{ker}(S))^+
 \end{gather}
 with inner product
 \begin{gather}\label{sip}
   \langle [v_1], [v_2] \rangle_{H^+} := \langle [v_2^+], [v_1^+] \rangle_H\;.
 \end{gather}
 We note that $S$ maps $H$
 onto the space of smooth solutions to the Dirac equation whose supports have compact intersections
 with any Cauchy surface.
 The construction of the Dirac field starts with the Hilbert space
 $\mathcal{H}:= \overline{H \oplus H^+}$ and
 the antiunitary involution
 \begin{gather}
  \mathcal{K}: \mathcal{H} \to \mathcal{H}, \quad u \oplus v \to v^+\oplus u^+\;.
 \end{gather}
 By construction we have a map
 \begin{gather*}
  \eta: \Gamma_0(DM \oplus DM^*) \to \Hilbert; \quad
  f \oplus g \to [f] \oplus [g]
 \end{gather*}
 with dense range.
 Let $G$ be the identity component of the group of isometries of $M$ and $\tilde G$
 its universal covering group. $\tilde G$ acts
 canonically on $FM$ and since $\tilde G$ is simply connected this actions
 lifts uniquely to a smooth
 action  on $SM$, yielding a smooth equivariant action on the
 bundles $DM$ and $DM^*$, making them $\tilde G$-vector bundles.
 The Dirac operator $\dirac$ and its conjugate
 $\dirac^+:={}^+ \circ \dirac \circ^+$ are both invariant under these actions
 and hence the representation of $\tilde G$ on $\Gamma_0(DM \oplus DM^*)$
 gives rise to a unitary representation $\rho$ of $\tilde G$ on $\Hilbert$. 
 Taking $E:=DM \oplus DM^*$ one shows that
 $(\Hilbert,\mathcal{K},E,\eta, \rho)$ is a linear fermionic
 field theory on $M$. Moreover, the differential operator
 $P=\dirac \oplus \dirac^+: \Gamma_0(E) \to \Gamma_0(E)$
 maps to the kernel of $\eta$, i.e. $\eta \circ P=0$.
 The square of $P$ has metric principal part. \vspace{0.3cm}\\
\bf The real scalar field \rm (see \cite{Dimock:1980hf}).
 The construction of the Klein-Gordon field starts with the Klein-Gordon operator
 for mass $m \geq 0$:
 \begin{gather}
   P:=\square_g +m^2,
 \end{gather}
 where $\square_g=g^{ik} \nabla_i \nabla_k\;$ 
 and $\nabla$ is the Levi-Civita covariant derivative.
 This operator acts on the real-valued smooth functions with compact support
 $C^\infty_{0r}(M)$.
 It has unique advanced and retarded fundamental solutions
 $F_s^\pm: C^\infty_{0r}(M) \to C^\infty_r(M)$
 satisfying
 $$ P F_s^\pm = F_s^\pm P = \mathrm{id} \quad \textrm{on}
 \quad C^\infty_{0r}(M)\;,$$
 $$ \textrm{supp}(F_s^\pm f) \subset J^\pm(\textrm{supp}(f))\;.$$
 With $F_s:=F_s^+-F_s^-$, $\hat\sigma(f_1,f_2):=\int_M f_1 F_s(f_2) w$ defines
 an antisymmetric bilinear form on $C^\infty_{0r}(M) \times C^\infty_{0r}(M)$, where $w$
 is the pseudo-Riemannian volume form on $M$.
 Defining $\mathcal{W}:= C^\infty_{0r}(M)/\textrm{ker}(F_s)$ with quotient map
 $\eta$, the bilinear form $\sigma(\eta(f_1),\eta(f_2)):=\hat \sigma(f_1,f_2)$ on
 $\mathcal{W}$ is symplectic.
 We have a canonical linear action of the group $G$ on $C^\infty_{0r}(M)$ which
 leaves $\textrm{ker}(F_s)$ and $\hat\sigma$ invariant and hence gives
 a representation $\rho$ of $G$ on $\mathcal{W}$ by symplectomorphisms.
 Taking the trivial bundle $M \times \rz$ for $E$, one shows that
 $(\mathcal{W},\sigma,E,\eta, \rho)$
 is a linear bosonic field theory on $M$ and $\eta \circ P=0$.
 Note that $P$ has metric principal part.

\begin{rem}
 Since the complex scalar field consists of two
 independent real scalar fields it fits into this framework as well.
\end{rem}\vspace{0.3cm}
\noindent
\bf The Proca field \rm (see \cite{Furlani:1999kq}).
 Let $d$ be the exterior derivative of differential forms,
 $*$ the Hodge star operator and $\delta=* d *$.
 Take the cotangent bundle for $E$.
 For mass $m > 0$ the Proca equation for sections
 $f \in\Gamma_0(E)$ is
 \begin{gather}
   (\delta \circ d +m^2) f =0,
 \end{gather}
 which is equivalent to the hyperbolic system
 \begin{gather}
   (\square_g + m^2)f=(\delta \circ d + d \circ \delta +m^2)f=0,\\
   \delta f =0.
 \end{gather}
 We define the differential operator $\tilde P:\Gamma_0(E) \to \Gamma_0(E)$ by
 \begin{gather}
   \tilde P:=\square_g + m^2.
 \end{gather}
 It has unique advanced and retarded fundamental solutions
 $\tilde F_p^\pm: \Gamma_0(E) \to \Gamma(E)$
 satisfying
 $$ \tilde P \tilde F_p^\pm = \tilde F_p^\pm \tilde P = \mathrm{id} \quad \textrm{on}
 \quad \Gamma_0(E)\;,$$
 $$ \textrm{supp}(\tilde F_p^\pm f) \subset J^\pm(\textrm{supp}(f))\;.$$
 We define the operators
 $F_p^\pm:=( m^{-2} d \circ \delta+1)\tilde F_p^\pm$.
 It is not difficult to see that the $F_p^\pm$ are the unique fundamental
 solutions for the operator $P = \delta \circ d +m^2$ with the above
 properties. With $F_p:=F_p^+-F_p^-$,
 $\hat\sigma(f_1,f_2):=\int_M f_1 \wedge * F_p(f_2)$ defines
 an antisymmetric bilinear form on $\Gamma_0(E) \times \Gamma_0(E)$.
 Taking $\mathcal{W}:= \Gamma_0(E)/\textrm{ker}(F_p)$ with quotient map
 $\eta$, the bilinear form $\sigma(\eta(f_1),\eta(f_2)):=\hat \sigma(f_1,f_2)$ on
 $\mathcal{W}$ is symplectic. The pullback of forms induces
 a linear action of $G$ on $\Gamma_0(E)$ which leaves $\textrm{ker}(F_p)$ and $\hat\sigma$
 invariant and hence gives rise to a representation $\rho$ of $G$ on $\mathcal{W}$
 by symplectomorphisms. Again one can show that
 $(\mathcal{W},\sigma,E,\eta, \rho)$ is a linear bosonic field theory
 and moreover, $\eta \circ \tilde P=\eta \circ P=0$.
 \vspace{0.4cm}\\
 One gets the following corollary.

\begin{cor}
 Let $(M,g,h_t)$ be a connected stationary globally hyperbolic oriented
 time-oriented 4-dimensional Lorentzian manifold.
 Let $\openset \to \feld(\openset)$  be a net of field
 algebras for one of the following free fields:
 \begin{itemize}
   \item The real or complex scalar field for mass $m \geq 0$,
   \item The Proca field for mass  $m > 0$,
   \item The Dirac field for mass $m \geq 0$,
 \end{itemize}
 Assume that $\omega$ is a state over the field algebra $\feld$ which we require to be quasifree
 and continuous in the bosonic case and which is a ground- or KMS-state with
 respect to the canonical time translations.
 Then $\omega$ has the Reeh-Schlieder property.
\end{cor}

\section{Acknowledgements}
The author would like to thank Prof. M. Wollenberg and Dr. R. Verch for
useful discussions and comments.
This work was supported by the
Deutsche Forschungsgemeinschaft within the scope of the postgraduate
scholarship programme ``Graduiertenkolleg Quantenfeldtheorie'' at the
University of Leipzig.

\end{document}